\documentclass{PoS}

\title{The NESSiE Concept for Sterile Neutrinos}

\ShortTitle{Nessie sterile neutrinos}

\author{{Luca Stanco}\thanks{For the NESSiE Collaboration.}
\\
        I.N.F.N. Padova, Via Marzolo, 8 I-35131 Padova, Italy\\
        E-mail: \email{luca.stanco@pd.infn.it}}

\abstract{
Neutrino physics is nowadays receiving more and more attention as a possible source of information for the long--standing problem
of new physics beyond the Standard Model. The recent measurement of the third mixing angle $\theta_{13}$ in the standard 
mixing oscillation scenario encourages us to pursue the still missing results on leptonic CP violation and absolute neutrino
masses. However, several puzzling measurements exist, which deserve an exhaustive evaluation.\\
The NESSiE Collaboration has been setup to undertake a definitive experiment to clarify the  muon disappearance measurements at small $L/E$,
which will be able to put severe constraints to any model with more than the three-standard neutrinos, or even to robustly measure 
the presence of a new kind of neutrino oscillation for the first time.
Within the context of the current CERN project, aimed to
revitalize the neutrino field in Europe, we will illustrate the achievements that can be obtained by a double muon--spectrometer system, 
with emphasis on the search for sterile neutrinos.
}

\FullConference{XV Workshop on Neutrino Telescopes,\\
		11-15 March 2013\\
		Venice, Italy}

\newcommand{\BARII}                   {1}
\newcommand{\BARIU}                 {2}
\newcommand{\BOLOGNAI}         {3}
\newcommand{\BOLOGNAU}       {4}
\newcommand{\FRASCATI}          {5}
\newcommand{\LECCEI}             {6}
\newcommand{\LECCEU}           {7}
\newcommand{\LECCEING}       {8}
\newcommand{\LEBEDEV}          {9}
\newcommand{\MSU}                 {10}
\newcommand{\PADOVAI}          {11}
\newcommand{\PADOVAU}        {12}
\newcommand{\ROMAU}            {13}
\newcommand{\ZAGREB}             {14}
\newcommand{\NessieInstitutes}{
\BARII        . INFN, Sezione di Bari, 70126 Bari, Italy \\
\BARIU        . Dipartimento di Fisica dell'Universit\`a  di Bari, 70126 Bari, Italy \\
\BOLOGNAI     . INFN, Sezione di Bologna, 40127 Bologna, Italy \\
\BOLOGNAU     . Dipartimento di Fisica dell'Universit\`a  di Bologna, 40127 Bologna, Italy \\
\FRASCATI    . Laboratori Nazionali di Frascati dell'INFN, 00044 Frascati (Roma), Italy \\
\LECCEI        . INFN, Sezione di Lecce, 73100 Lecce, Italy \\
\LECCEU        . Dipartimento di Matematica e Fisica dell'Universit\`a  del Salento, 73100 Lecce, Italy \\
\LECCEING        . Dipartimento di Ingegneria dell'Innovazione dell'Universit\`a  del Salento, 73100 Lecce, Italy \\
\LEBEDEV     . Lebedev Physical Institute of Russian Academy of Science, Leninskie pr., 53, 119333 Moscow, Russia.\\
\MSU       . Lomonosov Moscow State University (MSU SINP), 1(2) Leninskie gory, GSP-1, 119991 Moscow, Russia\\
\PADOVAI      . INFN, Sezione di Padova, 35131 Padova, Italy \\
\PADOVAU      . Dipartimento di Fisica e Astronomia dell'Universit\`a  di Padova, 35131 Padova, Italy \\
\ROMAU        . Dipartimento di Fisica dell'Universit\`a  di Roma ``La Sapienza" and INFN, 00185 Roma, Italy \\
\ZAGREB             . Rudjer Boskovic Institute, Bijenicka 54, 10002 Zagreb, Croatia\\
**~Now at CERN, CH-1211 Geneva 23, Switzerland\\
\ddag~Also at Centre de Recherche en Astronomie Astrophysique et Gophysique, Alger, Algeria\\
}
\newcommand{\NessieAuthorList}{
\noindent 
A.~Anokhina$^{\MSU}$,
A.~Bagulya$^{\LEBEDEV}$,
M.~Benettoni$^{\PADOVAI}$,
P.~Bernardini$^{\LECCEU, \LECCEI}$,
A.~Bertolin$^{\PADOVAI}$,
R.~Brugnera$^{\PADOVAU, \PADOVAI}$,
M.~Calabrese$^{\LECCEI}$,
A.~Cecchetti$^{\FRASCATI}$,
S.~Cecchini$^{\BOLOGNAI}$,
M.~Chernyavskiy$^{\LEBEDEV}$,
G.~Collazuol$^{\PADOVAU, \PADOVAI}$,
P.~Creti$^{\LECCEI}$,
F.~Dal~Corso$^{\PADOVAI}$,
O.~Dalkarov$^{\LEBEDEV}$,
A.~Del~Prete$^{\LECCEING}$,
I.~De~Mitri$^{\LECCEU, \LECCEI}$,
G.~De~Robertis$^{\BARII}$,
M.~De~Serio$^{\BARIU, \BARII}$,
L.~Degli~Esposti$^{\BOLOGNAI}$,
D.~Di~Ferdinando$^{\BOLOGNAI}$,
U.~Dore$^{\ROMAU}$,
S.~Dusini$^{\PADOVAI}$,
T.~Dzhatdoev$^{\MSU}$,
C.~Fanin$^{\PADOVAI}$,
R.~A.~Fini$^{\BARII}$,
G.~Fiore$^{\LECCEI}$,
G.~Galati$^{\BARIU}$,
A.~Garfagnini$^{\PADOVAU, \PADOVAI}$,
G.~Giacomelli$^{\BOLOGNAU, \BOLOGNAI}$,
R.~Giacomelli$^{\BOLOGNAI}$,
S.~Golovanov$^{\LEBEDEV}$,
C.~Guandalini$^{\BOLOGNAI}$,
M.~Guerzoni$^{\BOLOGNAI}$,
B.~Klicek$^{\ZAGREB}$,
U.~Kose$^{\PADOVAI *}$,
K.~Jakovcic$^{\ZAGREB}$,
G.~Laurenti$^{\BOLOGNAI}$,
M.~Laveder$^{\PADOVAU, \PADOVAI}$,
I.~Lippi$^{\PADOVAI}$,
F.~Loddo$^{\BARII}$,
A.~Longhin$^{\FRASCATI}$,
P.~Loverre$^{\ROMAU}$,
M.~Malenica$^{\ZAGREB}$,
G.~Mancarella$^{\LECCEU, \LECCEI}$,
G.~Mandrioli$^{\BOLOGNAI}$,
A.~Margiotta$^{\BOLOGNAU, \BOLOGNAI}$,
G.~Marsella$^{\LECCEU, \LECCEI}$,
N.~Mauri$^{\FRASCATI}$,
E.~Medinaceli$^{\PADOVAU, \PADOVAI}$,
A.~Mengucci$^{\FRASCATI}$,
M.~Mezzetto$^{\PADOVAI}$,
R.~Michinelli$^{\BOLOGNAI}$,
R.~Mingazheva$^{\LEBEDEV}$,
O.~Morgunova$^{\MSU}$,
M.~T.~Muciaccia$^{\BARIU, \BARII}$,
D.~Orecchini$^{\FRASCATI}$,
A.~Paoloni$^{\FRASCATI}$,
G.~Papadia$^{\LECCEING}$,
L.~Paparella$^{\BARIU, \BARII}$,
A.~Pastore$^{\BARII}$,
L.~Patrizii$^{\BOLOGNAI}$,
N.~Polukhina$^{\LEBEDEV}$,
M.~Pozzato$^{\BOLOGNAU, \BOLOGNAI}$,
M.~Roda$^{\PADOVAU, \PADOVAI}$,
T.~Roganova$^{\MSU}$,
G.~Rosa$^{\ROMAU}$,
Z.~Sahnoun$^{\BOLOGNAI \ddag}$,
S.~Simone$^{\BARIU, \BARII}$,
M.~Sioli$^{\BOLOGNAU, \BOLOGNAI}$,
C.~Sirignano$^{\PADOVAU, \PADOVAI}$,
G.~Sirri$^{\BOLOGNAI}$,
M.~Spurio$^{\BOLOGNAU, \BOLOGNAI}$,
L.~Stanco$^{\PADOVAI, a}$,
N.~Starkov$^{\LEBEDEV}$,
M.~Stipcevic$^{\ZAGREB}$,
A.~Surdo$^{\LECCEI}$,
M.~Tenti$^{\BOLOGNAU, \BOLOGNAI}$,
V.~Togo$^{\BOLOGNAI}$,
M.~Ventura$^{\FRASCATI}$, 
M.~Vladymyrov$^{\LEBEDEV}$ and
M.~Zago$^{\PADOVAI}$.\\
{\em (a)} Spokesperson
}

\begin{document}

\centerline{\bf The NESSiE Collaborations}

{\noindent \\ \NessieAuthorList }

\begin{flushleft}
\footnotesize\em{\NessieInstitutes }
\end{flushleft}

\newpage

\section{Introduction}
The unfolding of the physics of the neutrino is a long and exciting story spanning the last 80 years. Over this time the interchange of 
theoretical hypotheses and experimental facts has been one of the most fruitful demonstrations of the progress of knowledge in physics.
The work of the last decade and a half finally brought a coherent picture within the Standard Model (SM) (or some small extensions of it),
namely the mixing of three neutrino flavour states with three  $\nu_1$, $\nu_2$ and $\nu_3$ mass eigenstates. 
The last unknown mixing angle, $\theta_{13}$, was recently measured~\cite{theta13} but still
many questions remain unanswered to completely settle the scenario: the absolute masses, 
the Majorana/Dirac nature and the existence and magnitude of leptonic CP violation.
Answers to these questions will beautifully complete the (standard) three--neutrino model but they will hardly provide an insight into new physics 
Beyond the Standard Model (BSM). 
Many relevant questions will stay open: the reason for neutrinos, the relation between the
leptonic and hadronic sectors of the SM, the origin of Dark Matter and, overall, where and how to look for BSM physics.
Neutrinos may be an excellent source of BSM physics and their story is supporting that at length.

There are actually several experimental hints for deviations from the ``coherent'' picture described above.
Many unexpected results, not statistically significant on a single basis, appeared also in the last decade and a half,
bringing attention to the hypothesis of {\em sterile neutrinos}~\cite{pontecorvo}. A recent White Paper~\cite{whitepaper} contains a comprehensive
review of these issues. Here we focus on one of the most intriguing and long--standing unresolved result:
the unexpected oscillation of neutrinos at relatively small values of the ratio $L/E$ (distance in km, energy in GeV), corresponding to a scale
of $\mathcal{O}$(1) eV$^2$, incompatible 
with the much smaller values related to the atmospheric $|\Delta m_{32}^2|\simeq 2.4 \times 10^{-3}$~eV$^2$ and to the solar 
$\Delta m_{21}^2 \simeq 8 \times 10^{-5}$ eV$^2$ scales.

The first unexpected measurement came from an excess of $\overline\nu_e$ originating from an initial $\overline\nu_\mu$ beam from
Decay At Rest (LNSD~\cite{lsnd}). The LSND experiment saw a 3.8 $\sigma$ effect.  The subsequent  experiment with $\nu_{\mu}$ ($\overline\nu_\mu$) 
beam from accelerator, 
MiniBooNE~\cite{larnessie_5}, although confirming an independent 3.8 $\sigma$ effect after sustained experimental work,
was unable to draw conclusive results on the origin of the LSND effect having observed an excess at higher $L/E$ in an energy region where background
is high.

In recent years many phenomenological studies were performed by analyzing the LSND effect together with similar
unexpected results  coming from the measurement of lower than expected rates  of $\overline{\nu}_e$ and $\nu_e$ interactions
({\em disappearance}), either  from {\bf (a)} near-by nuclear reactors ($\overline\nu_e$)~\cite{reattori} or {\bf (b)} from Mega-Curie K-capture calibration
sources in the solar--$\nu_e$ Gallium experiments~\cite{larnessie_4}. These $\nu_e$ ($\overline\nu_e$) disappearance measurements, all at the statistical
level of 3-4 $\sigma$, could also be interpreted~\cite{giunti-laveder} as oscillations between neutrinos at large $\Delta m^2\simeq 1$~eV$^2$.
Several attempts were then tried to reach a coherent picture in terms of mixing between active and
sterile neutrinos, in $3+1$ and $3+2$~\cite{larnessie-6}  or even $3+1+1$~\cite{treunouno} or $3+3$~\cite{tretre} models, 
as extensions of the standard three--neutrino model.
We refer to~\cite{models,models1,giuntinew} as the most recent and industrious works where a very crucial issue is raised: 
``{\em a consistent interpretation of the global data in terms of neutrino oscillations is challenged by
the non-observation of a positive signal in $\nu_{\mu}$ disappearance experiments}''~\cite{models1}. In fact, since
the appearance probability of $\nu_e$ from $\nu_{\mu}$, in the two-flavour limit, is given by:
$$ P(\nu_{\mu}\to\nu_e)_{SBL}^{3+1} = 4 \vert U_{\mu 4}\vert^2 \vert U_{e 4} \vert^2  \sin^2 \frac{\Delta m^2_{41} L}{4E},$$
dependence on both the mixing elements with the sterile state, $\vert U_{e 4}\vert$ and $\vert U_{\mu 4}\vert$,
rises up.
Therefore, the presence of additional sterile states introduces quite naturally appearance and disappearance phenomena involving the flavour states in
all production channels. In particular, a $\nu_\mu$ disappearance effect has to be present and possibly
measured. It turns out that only old experiments and measurements are available for Charged Current (CC) $\nu_\mu$  
interactions at small $L/E$~\cite{CDHS}. The CDHS experiment reported in 1984 the non--observation of $\nu_\mu$ oscillations 
in the $\Delta m^2$ range $0.3$ eV$^2$ to $90$ eV$^2$. Their analyzed region of oscillation did not span however low values 
of the mixing parameter down to around $0.1$ in $\sin^2(2\theta)$. More recent results are available on $\nu_\mu$ disappearance
from MiniBooNE~\cite{mini-mu}, a joint MiniBooNE/SciBooNE analysis~\cite{mini-sci-mu1, mini-sci-mu2} and the Long--Baseline MINOS 
experiment~\cite{minos}. These results slightly extend the $\nu_\mu$ disappearance exclusion region, however
still leaving out the small--mixing region. Similar additional constraints on $\nu_{\mu}$ disappearance could possibly come
from the analysis of atmospheric neutrinos in IceCube~\cite{icecube}.

Despite this set of measurements being rather unsatisfactory when compared with the corresponding LSND allowed region 
that lies at somewhat lower values of the mixing angle, they are still sufficient to introduce 
tensions in all the phenomenological models developed so far~(see e.g. \cite{whitepaper,models,models1,giuntinew,reviews} for comprehensive and recent reviews).
Therefore it is mandatory to setup a new experiment able to improve
 the small--mixing angle region exclusion by at least one order of magnitude with respect to the current results. 
 In such a way one could also rule out the idea that the mixing angle extracted from LSND is larger than the true value due to a data over--fluctuation.
 Once again, the $\nu_\mu$ disappearance 
channel should be the optimal one to perform a full disentangling of the mechanism given the strong tension between the
$\nu_e$ appearance and $\nu_\mu$ disappearance around $\Delta m^2\simeq 1$ eV$^2$.
In fact, whereas the LSND effect may be confirmed by a more accurate $\nu_e$ oscillation measurement, only the  presence of a $\nu_\mu$ 
oscillation pattern could shed more light on the nature and the interpretation of the effect.

A very recent paper by us~\cite{stanco} depicted the phenomenology of the analysis that can be performed with a NESSiE-like experiment
in the newly proposed Neutrino Platform at CERN, in case a neutrino--beam will be available. 
An extension of more than one order of magnitude in the sensitivity to the mixing parameter will be reached.
This report, although corresponding to the presentation given at the Neutrino-Telescope Workshop in March 2013, is largely 
based on the more recent work of ~\cite{stanco} and the up-to-date development of the NESSiE experiment. In particular, the re-use 
of available hardware from the OPERA experiment is reported.

\section{The CERN experimental proposal}

The need for a definitive clarification on the possible existence of a neutrino mass scale around 1 eV has brought up several
proposals and experimental suggestions exploiting the sterile neutrino option by using different interaction channels and refurbished detectors.
In the light of the considerations discussed in the previous section there are essentially two sets of experiments which must be redone:
a)~the measurement of $\overline\nu_{e}$ neutrino fluxes at reactors (primarily the ``ILL'' one~\cite{ill}) together with refined and detailed computations
of the flux simulations~\cite{huber}; b)~the appearance/disappearance oscillation measurements at low $L/E$ with a standard muon neutrino beam with its intrinsic 
electron neutrino component.

There is actually another interesting option which comes from the Neutrino Factory studies and the very recently submitted 
EOI from $\nu$STORM~\cite{nustorm}. We refer to~\cite{winter} for a comprehensive review of the corresponding possible $\nu_e$ and $\nu_{\mu}$ 
disappearance effects. It is interesting to note that the figures of merit about $\nu_{\mu}$ disappearance as obtained in~\cite{stanco} are rather similar to or even slightly more
competitive than those illustrated in~\cite{winter} (e.g compare the exclusion regions in Fig.~6 of \cite{winter} and those in Fig.~\ref{ster-5} of~\cite{stanco} 
despite the use of different C.L.), not
forgetting the rather long time needed to setup the $\nu$STORM project.

Coming to experiments with standard beams, investigations are underway at CERN where two Physics Proposals~\cite{nessie, icarus} were submitted 
in October 2011 and later merged into a single Technical Proposal 
(ICARUS-NESSiE,~\cite{larnessie}). CERN has subsequently set up working groups for the 
proton beam extraction from the SPS, the secondary beam line and the needed infrastructure/buildings for the detectors. The work was reported in a 
recent LOI and at this Conference~\cite{edms}. 

The experiment is based on two Liquid Argon (LAr)--Time Projection Chambers (TPC)~\cite{icarus} of identical geometry (but different sizes) complemented by magnetized 
spectrometers~\cite{nessie} detecting electron and muon neutrino events at far and near
positions, 1600~m and 460 m away from the proton target, respectively.
The project will exploit the ICARUS T600 detector, the largest 
LAr--TPC ever built of about 600 ton mass, now presently in the LNGS underground 
laboratory where it was exposed to the CNGS beam. It is supposed to be moved at the CERN ``far'' position. 
An additional 1/4 of the T600 detector (T150) would be constructed from scratch as a clone of the original one, except for the dimensions,
 and located in the near site. Two spectrometers would be placed 
downstream of the two LAr--TPC detectors to greatly enhance the physics reach.
The spectrometers will exploit a bipolar magnet with instrumented iron slabs, and a newly designed
air--core magnet, to perform charge identification and muon momentum measurements in an extended energy range (from 0.5 GeV
or less to 10 GeV), over a transverse area larger than 50 m$^2$.

While the LAr--TPCs will mainly perform a direct measurement of electron neutrinos~\cite{rubbia} the spectrometers will
allow an extended exploitation of the muon neutrino component, with neutrino/antineutrino discrimination on an
event-by-event basis~\cite{stanco}.  

\subsection{The neutrino beam}
The proposed new neutrino beam will be constructed in the SPS North Area~\cite{edms}. The setup is based on a 100 GeV proton beam with a fast extraction 
scheme providing about 3.5~$\cdot$~10$^{13}$ protons/pulse, in two pulses of 10.5~$\mu$s durations\footnote{Pulses of 10.5~$\mu$s duration are normally  
put in coincidence
with the fast response of the spectrometers' detectors and efficiently used to reject the cosmic ray background. See e.g. ~\cite{opera-time} where a 
time resolution of less than 2 ns is reported for the detectors used in the OPERA experiment.} 
separated by 50~ms, that corresponds to about 4.5~$\cdot$~10$^{19}$ protons on target (p.o.t.) per year.
A target station will be located next to the so called  TCC2 target zone, 11~m underground, followed by a cylindrical He-filled decay pipe with a
length of about 110~m and a diameter of 3~m. The beam dump of 15~m in length, will be composed of iron blocks with a graphite inner
core. Downstream of the beam dump a set of muon chambers stations will act as beam monitors. The beam will point upward, with a slope of
about 5~mrad, resulting in a depth of 3 m for the detectors in the far site.

The current design of the focusing optics includes a pair of pulsed magnetic horns operated at relatively low currents. A graphite target
of about 1~m in length is deeply inserted into the first horn allowing a large acceptance for the focusing of low momentum pions emitted at
large angles. This design allows production of a spectrum peaking at about 2~GeV thus matching the most interesting domain of $\Delta
m^2$ with the detector locations at 460 and 1600~m from the target.

\begin{figure}[htbp]
\begin{center}
  \includegraphics[width=0.53\textwidth]{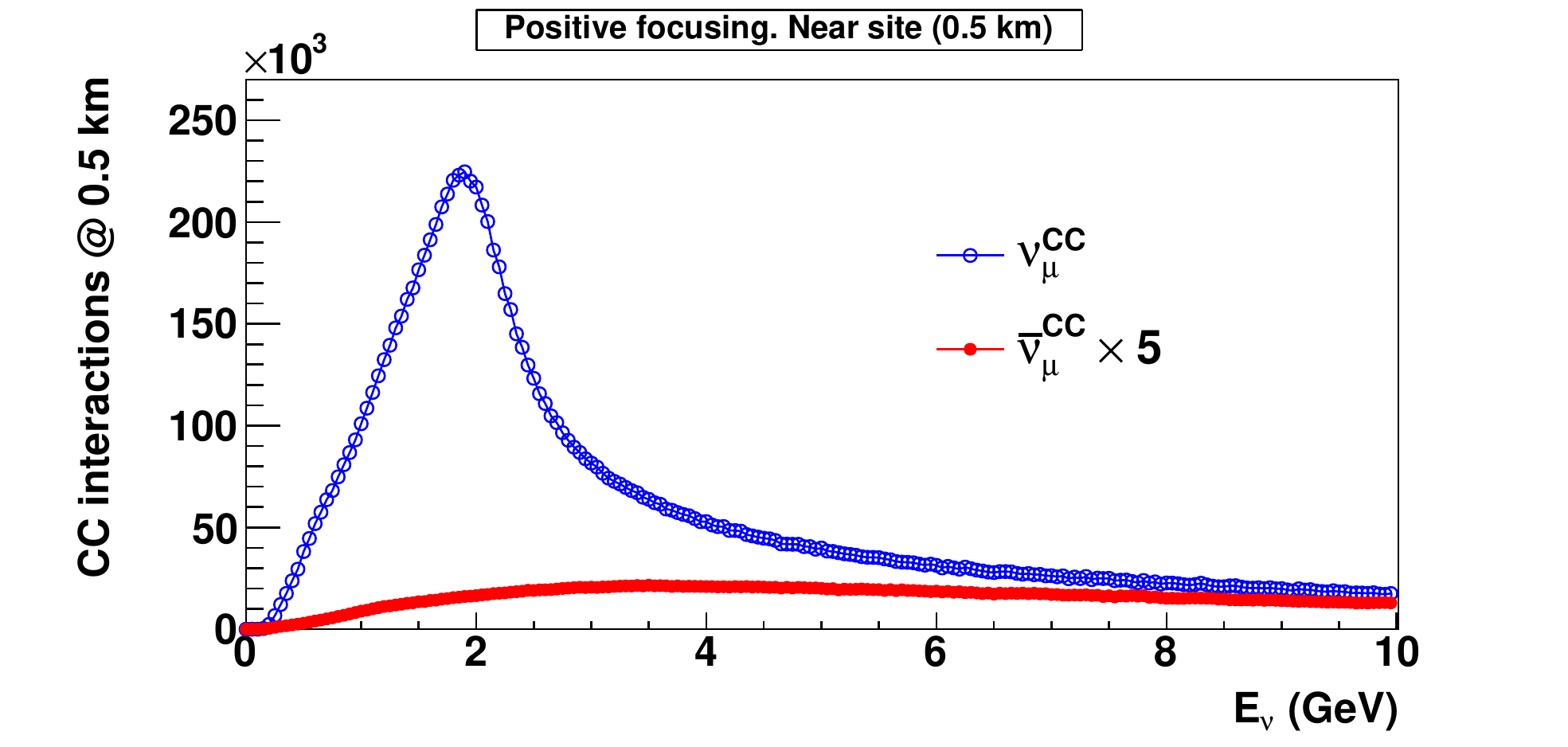}%
    \includegraphics[width=0.53\textwidth]{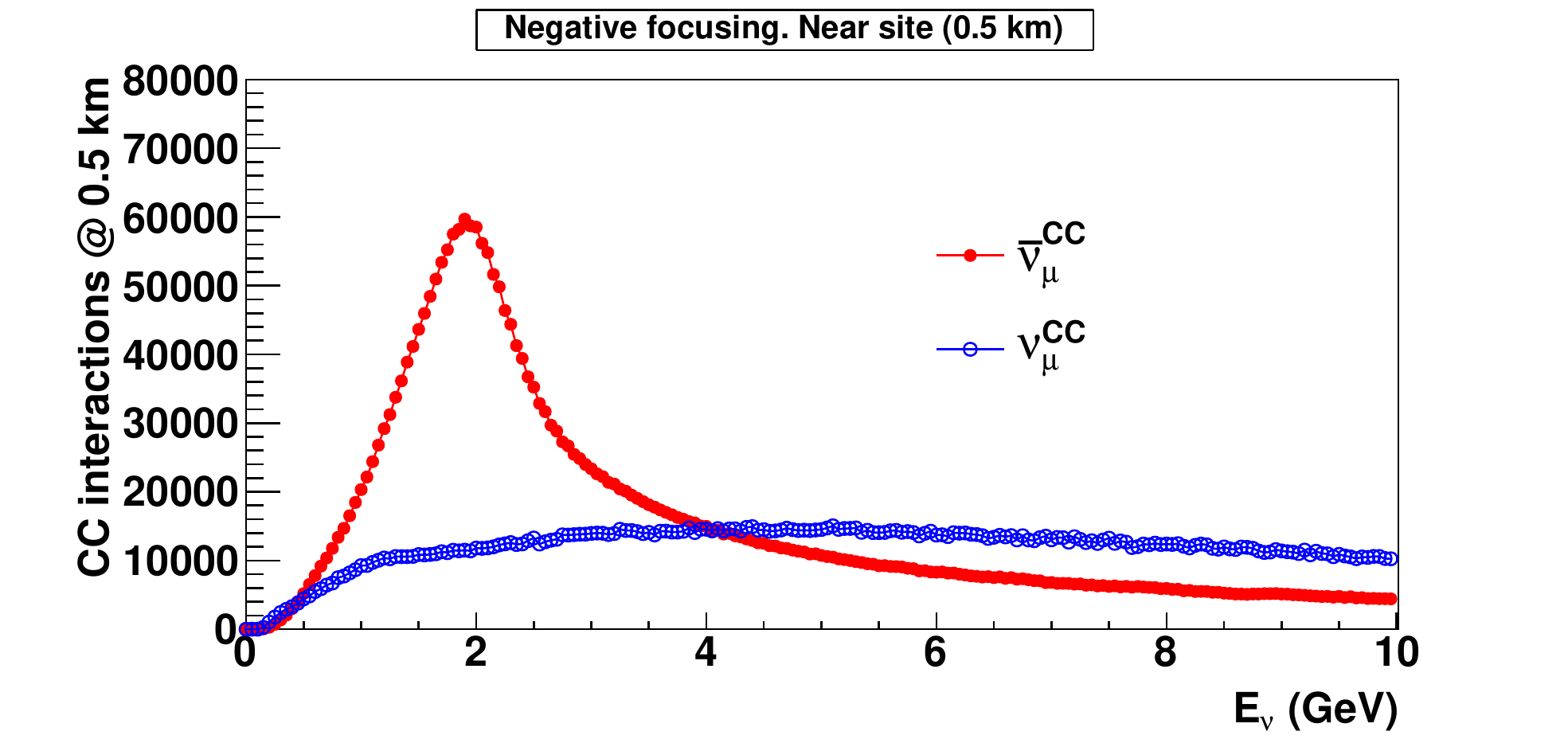}
  \includegraphics[width=0.53\textwidth]{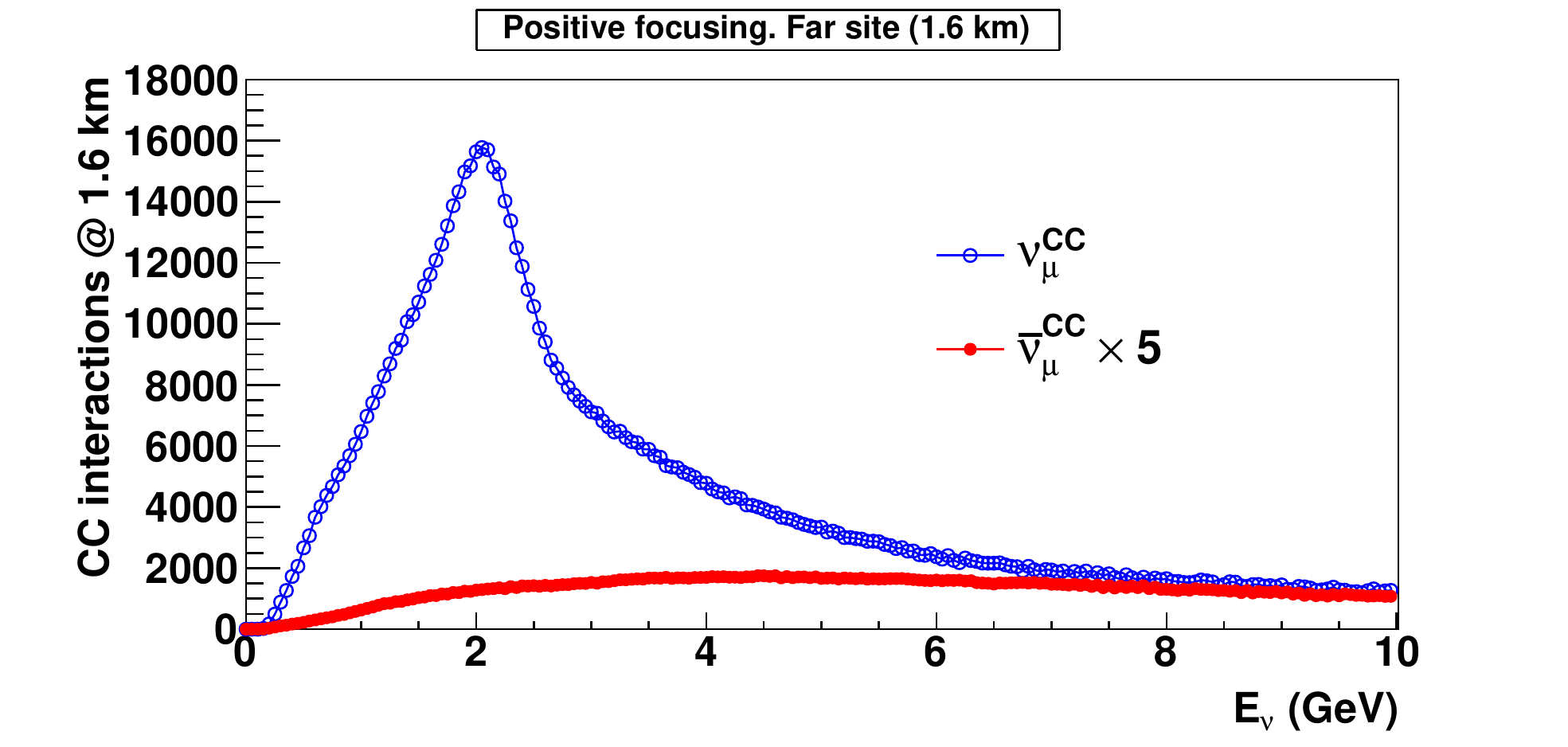}%
    \includegraphics[width=0.53\textwidth]{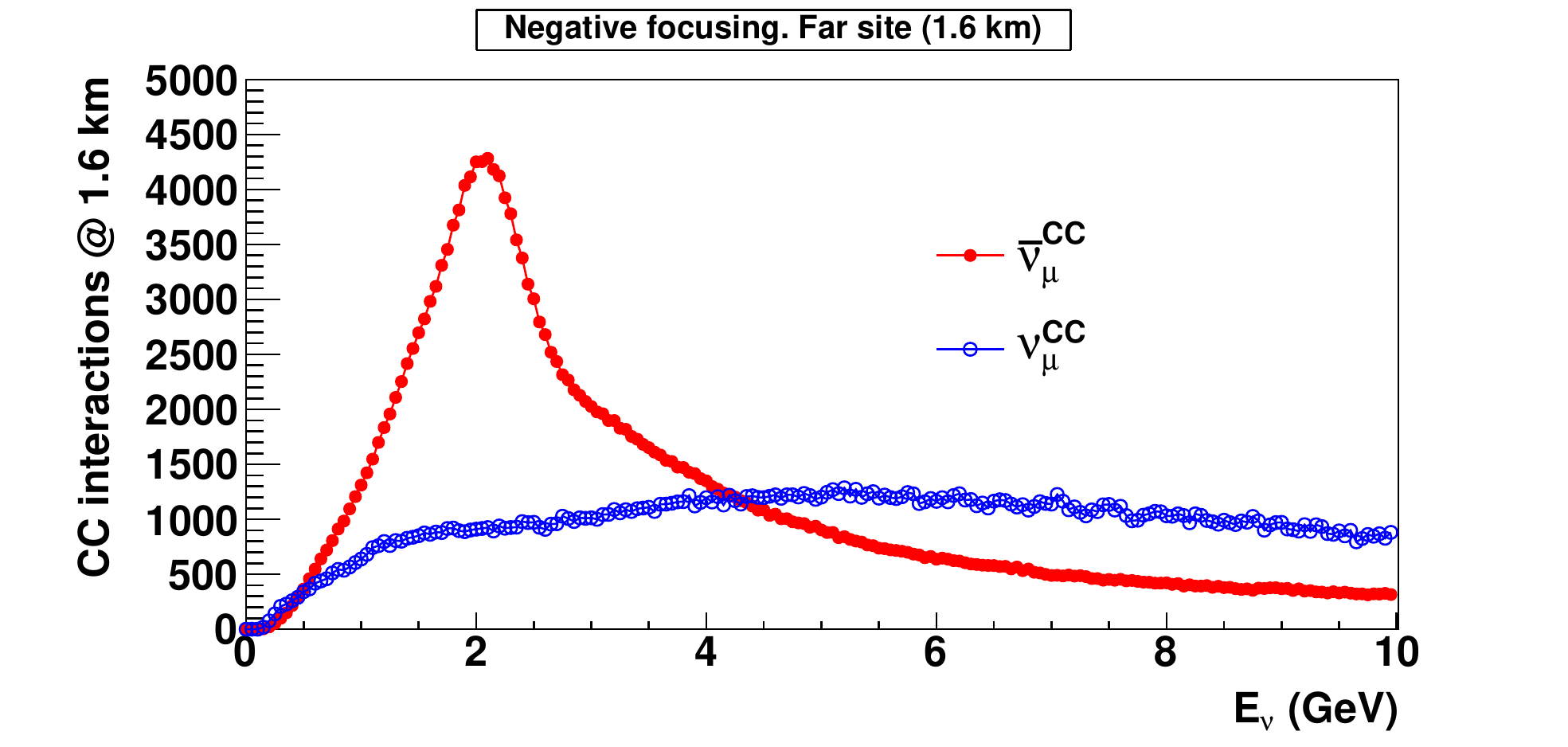}
    \caption{Expected neutrino CC interactions in the no--oscillation hypothesis for positive polarity (left) and negative polarity (right)
    for the new proposed CERN neutrino beam (elaborated from~\cite{edms}) and for an integrated luminosity of 1 year. 
    In the first (second) row the rates are shown at the near (far) position. 
     Note that the distributions of the antineutrino rates in case of positive polarity are multiplied by a factor 5 to allow a better visual inspection.}
    \label{beam-AL}
\end{center}
\end{figure}

The charged current event rates for $\nu_\mu$ and $\bar{\nu}_\mu$ at the near and far detectors are shown in Fig.~\ref{beam-AL} for the positive and
negative focusing configuration.

A relevant contamination of $\nu_\mu$ in the negative polarity configuration is visible especially at high energy. This
component arises as a result of the decays of high energy poorly de--focused mesons produced at small angles. 
The charge discrimination of the magnetic system described below will allow an efficient discrimination of these two
components with a charge confusion below or of the order of 1\% from sub--GeV (0.3--0.5 GeV) up to momenta around 8--10 GeV~\cite{nessie}.

\section{Spectrometer requirements}

The main purpose of the spectrometers placed downstream of the LAr--TPC is to provide charge identification and momentum 
reconstruction for the muons produced in neutrino interactions occurring in the LAr volume or in the magnetized iron of the spectrometers. 
In order to perform this measurement with high precision and in a wide energy range, from sub-GeV to multi-GeV, a massive iron-core dipole magnet 
(ICM) is coupled to an air--core magnet (ACM) in front of it~\cite{nessie}. Low momentum muons will be measured by the ACM while the ICM will be 
employed at higher momenta. 

\begin{figure}[htbp]
\begin{center}
  \includegraphics[width=0.7\textwidth]{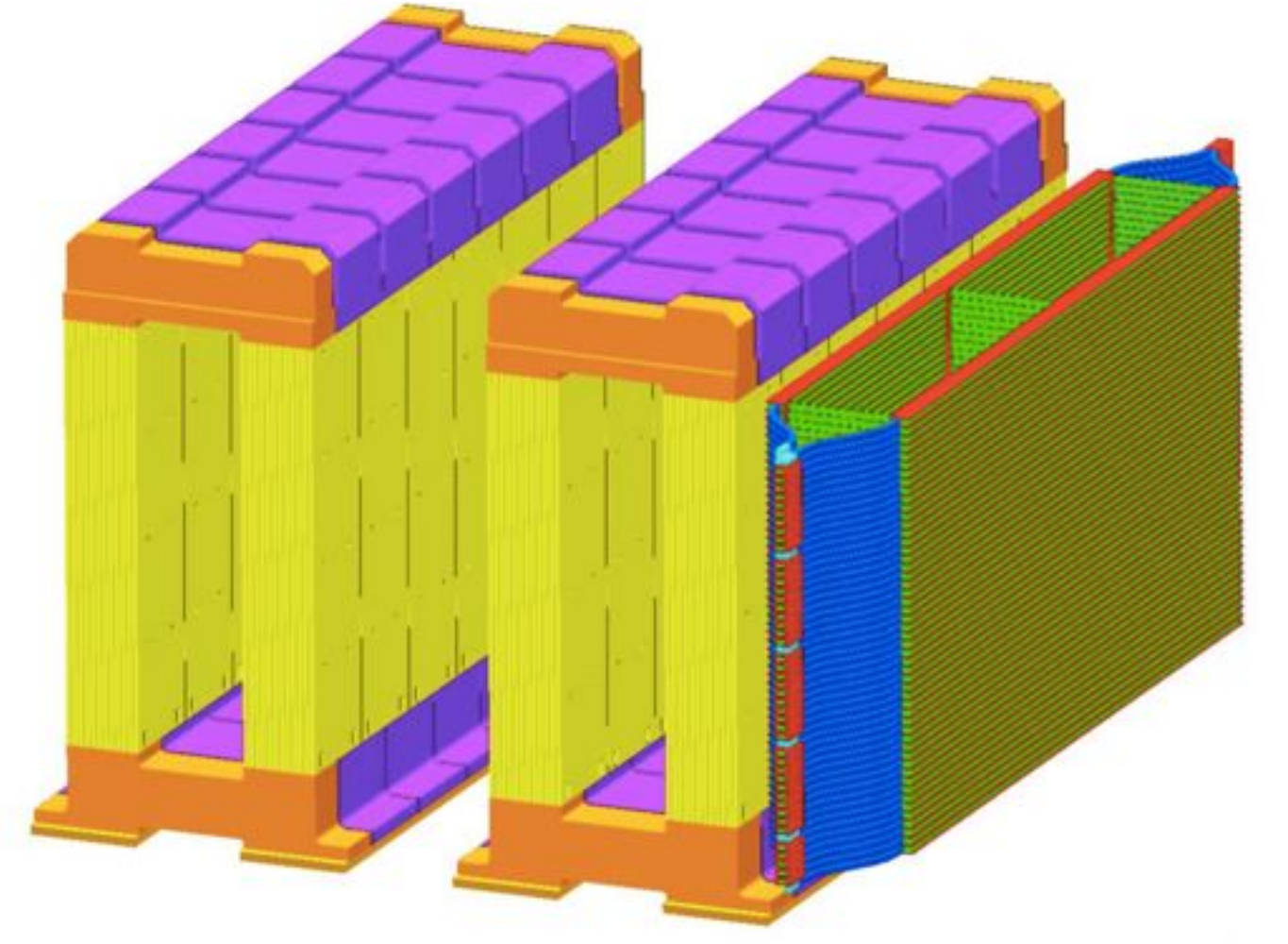}
    \caption{Sketch of the far site spectrometer that could be built by extensively reusing materials available from the OPERA 
     spectrometers~\cite{opera}. The basic elements of the new ACM concept are also depicted.
     Neutrinos are traveling from right to left.}
    \label{nessie-far}
\end{center}
\end{figure}

As considered in the previous sections the definition of two sites, near and far, constitutes a fundamental issue for each physics program
which aims to perform any sterile neutrino search. The two layouts have to be as similar as possible in order to allow an
almost complete cancellation of the systematic uncertainties when comparing the measurements made at the near and far sites. 
Hence the near spectrometer will be an exact clone of the far one, with identical thickness along the beam axis but a scaled transverse size.
A sketch of the possible far site NESSiE detector is shown in Fig.~\ref{nessie-far}.

The key feature of the ACM is the large geometric and momentum acceptance. The need of a low momentum threshold for muon reconstruction can be
met using a magnetized air--volume. The only dead material along the muon path is given by the conductors needed to generate the magnetic field and the
position detectors instrumenting the magnet itself. For the conductors the use of aluminum instead of copper is preferable
due to the lower $Z$ and a density lower by a factor 3. The magnetic field needed in the low momentum
range covered by the ACM is in the range 0.1 - 0.15 T. A spatial resolution in the range of 0.1 - 1 mm can be reached using drift tubes~\cite{opera-hpt} as
high precision trackers in combination with scintillator strip detectors~\cite{opera-scint}. These could provide the external trigger 
needed by the drift tubes and a coarse spatial measurement in the non--bending direction i.e. the direction parallel to the drift tubes. Silicon 
photomultiplier (SiPM) devices may eventually be used to read out some scintillator planes embedded in the magnetic field.

The general layout of the two OPERA~\cite{opera} iron spectrometers fulfills the requirements of the ICM detector, and
they could eventually be used for the CERN project putting the two magnets one after the other in order to obtain a total of 2.5 m longitudinal thickness of iron.

The OPERA spectrometers are built assembling vertical iron plates (slabs) in a planar structure of 875 cm (width) $\times$ 800 cm (height). Each 
passive plane is made out of seven adjacent iron slabs. Resistive Plate Chambers (RPC)~\cite{opera-rpc} are sandwiched between iron planes; 
21 RPC detectors, 
arranged in seven rows and three columns, are used in each active plane. Each RPC detector has a rectangular shape and covers an area of about
3.2 m$^2$. The RPCs provide tracking measurements with about 1 cm resolution using the digital read-out of strips with a 2.6 cm pitch in the bending direction
and 3.5 cm in the non--bending direction. The magnet is made of two arms with 22 RPC layers alternating with iron
layers. Top and bottom iron yokes are connecting the arms. Copper coils surrounding the yokes are used to generate a magnetic field of about 1.5 T in the iron 
circuit. 
A total of 924 RPC chambers  are needed to instrument each spectrometer. 
They could  be recovered from the OPERA spectrometers and re-used. 
These RPC chambers are standard 2 mm gap chambers with bakelite electrodes with resistivity in the range from 
10$^{11}$ to 5$\times$10$^{12}$ $\Omega$ $\cdot$ cm at $T=20 ^{\circ}$C. In OPERA they are operated in the streamer regime with a gas mixture made of 
$Ar/C_2H_2F_4/I$-$C_4H_{10}/SF_6$ in the volume ratios of 75.4/20/4/0.6 at five refills/day. The high-amplitude streamer signals produced by charged tracks crossing the
gas volume allow to house the Front-End (FE) discriminators in racks placed on top of the spectrometer. A totally new FE electronic for a surface
and closeness to the neutrino production, i.e. at a very larger rate than that available at LNGS, has been already developed by the NESSiE Collaboration.

\begin{figure}[htbp]
\begin{center}
  \includegraphics[width=0.43\textwidth]{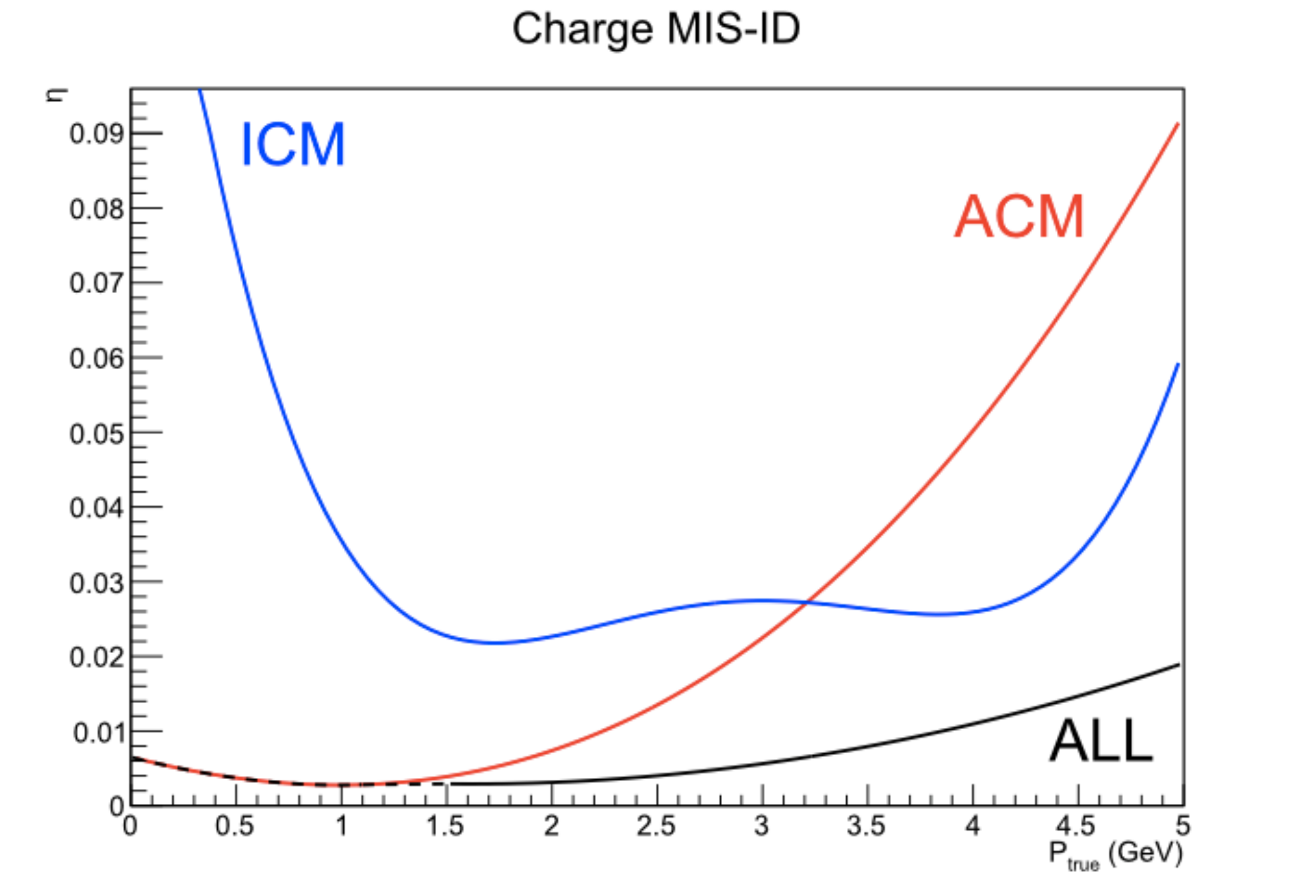}%
    \includegraphics[width=0.43\textwidth]{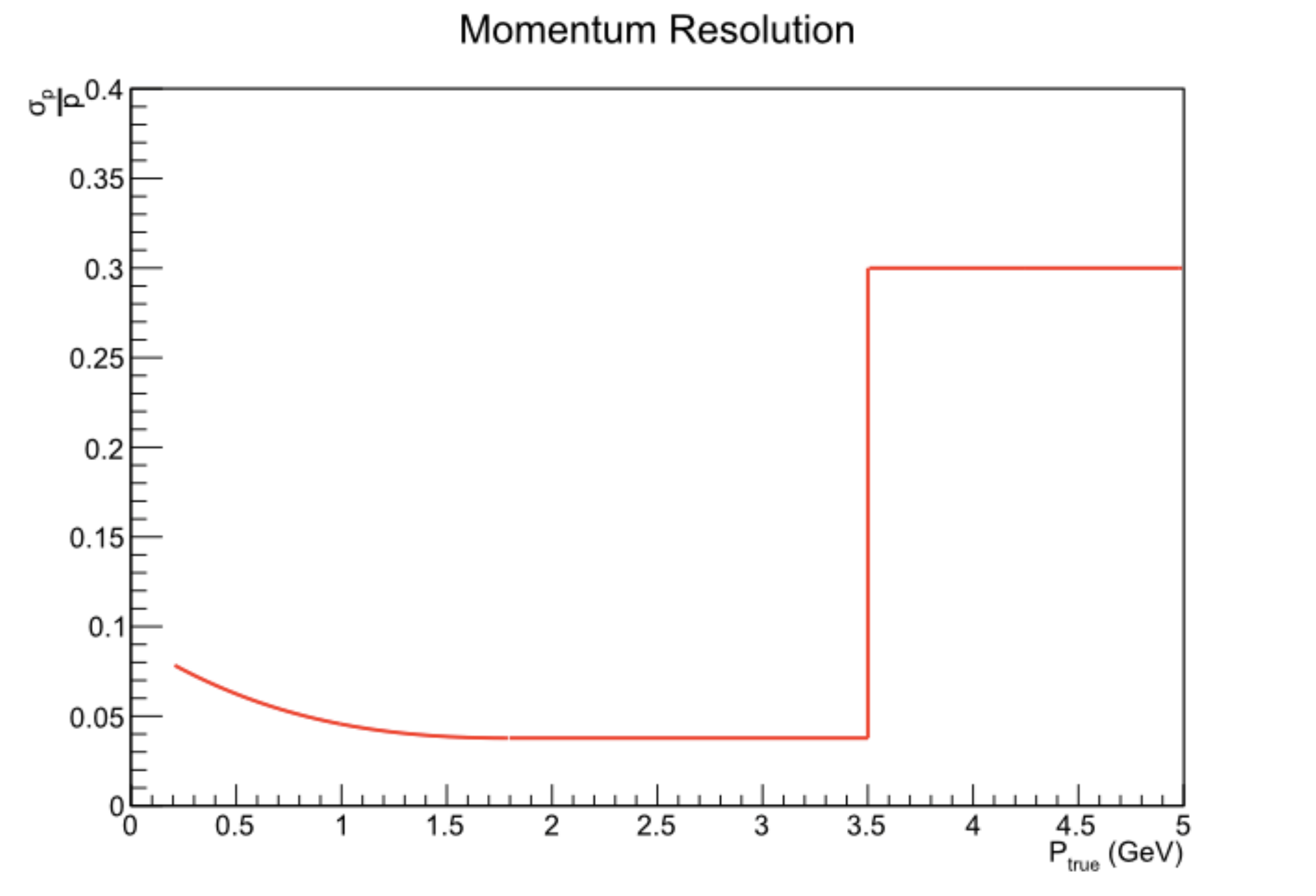}
    \caption{Sensitivity on the charge-ID and the momentum resolution for muon tracks coming from CC interactions. The full reconstruction procedure has been applied.
     On the left figure, the blue line corresponds to the mis-ID due only to the ICM part, while the red line is for the ACM part. The combination of the two magnetic system
     allow to achieve a mis-ID below 2\% up to 5 GeV in the muon momentum.
     The right figure shows the relative resolution on the muon momentum. There are two regions: the first up to 3.5 GeV with a very good resolution of about 5\%
     and the second region that corresponds to muons with momentum greater than 3.5 GeV. The first region corresponds to the momentum measured by range by the ICM.
     The second region is the resolution obtained by analyzing the curvature either in the ICM or the ACM (similar response by the two system is observed).}
    \label{muons-id}
\end{center}
\end{figure}

\section{Perspectives}

The magnetic detector system that has to be developed for the $\nu_{\mu}$ disappearance measurement should take into account all the 
considerations depicted in previous sections. The system developed by the NESSiE Collaboration~\cite{nessie} is actually well suited since it 
couples a very powerful high-$Z$ magnet for the momentum measurement via {range} to a low-$Z$ magnet, to extend the useful muon momentum interval as low as possible,
to allow charge discrimination on an event--by--event basis and to allow NC event measurement whether coupled to an adequate (but not necessarily
highly performant and with large mass) detector to identify NC events.
The sensitivities that can be obtained are depicted in Fig.~\ref{muons-id} for the charge mis-identification and the momentum resolution of the muon track.
Full simulation and reconstruction of the neutrino-interaction events have been applied.

For one year of operation at the CERN beam, either with negative or positive polarity beam, 
Table~\ref{larnessie_tab1} reports the expected interaction rates in the LAr--TPCs at the Near (fiducial 119 t) and Far locations (fiducial 476 t), 
and the expected rates of fully reconstructed events in the NESSiE spectrometers at the Near (fiducial 241 t) and Far locations 
(fiducial 661 t), with and without LAr contribution. Significant results will be achievable. Both $\nu_e$ and $\nu_{\mu}$ disappearance modes will be used, 
in addition to $\nu_e$ appearance mode, to add conclusive information on the sterile mixing angles, either in the $3 + 1$ or the $3 + 2$ scenarios.

\begin{table}[htbp]
\footnotesize
  \begin{center}
  \begin{tabular}{lcccc}
  \hline
  & NEAR                            & NEAR             & FAR                        & FAR            \\
  &       (Neg. foc.)           & (Pos. foc.)  & (Neg.)            & (Pos.) \\
  \hline
 $\nu_e + \overline{\nu}_e$ (LAr) & 35 K & 54 K & 4.2 K & 6.4 K \\
 $\nu_{\mu} +  \overline{\nu}_{\mu}$ (LAr) & 2000 K & 5200 K & 270 K & 670 K \\ 
 App. Test Point & 590 & 1900 & 360 & 910 \\
                                            &        &         &      &       \\
 $\nu_{\mu}$ CC (NESSiE$+$LAr) & 230 K & 1200 K & 21 K & 110 K \\
  $\nu_{\mu}$ CC (NESSiE alone) & 1150 K & 3600 K & 94 K & 280 K \\
  $\overline{\nu}_{\mu}$ CC (NESSiE$+$LAr) & 370 K & 56 K & 33 K & 6.9 K \\
  $\overline{\nu}_{\mu}$ CC (NESSiE alone) & 1100 K & 300 K & 89 K & 22 K \\
 Disapp. Test Point & 1800 & 4700 & 1700 & 5000 \\
 \hline
  \end{tabular}
  \caption{The expected rates of interaction (LAr) and reconstructed (NESSiE) events for 1 year of operation.Values for 
  $\Delta m^2$ around 2 eV$^2$ are reported as example.}
  \label{larnessie_tab1}
  \end{center}
\end{table}

\begin{figure}[htbp]
\begin{center}
  \includegraphics[width=0.35\textwidth]{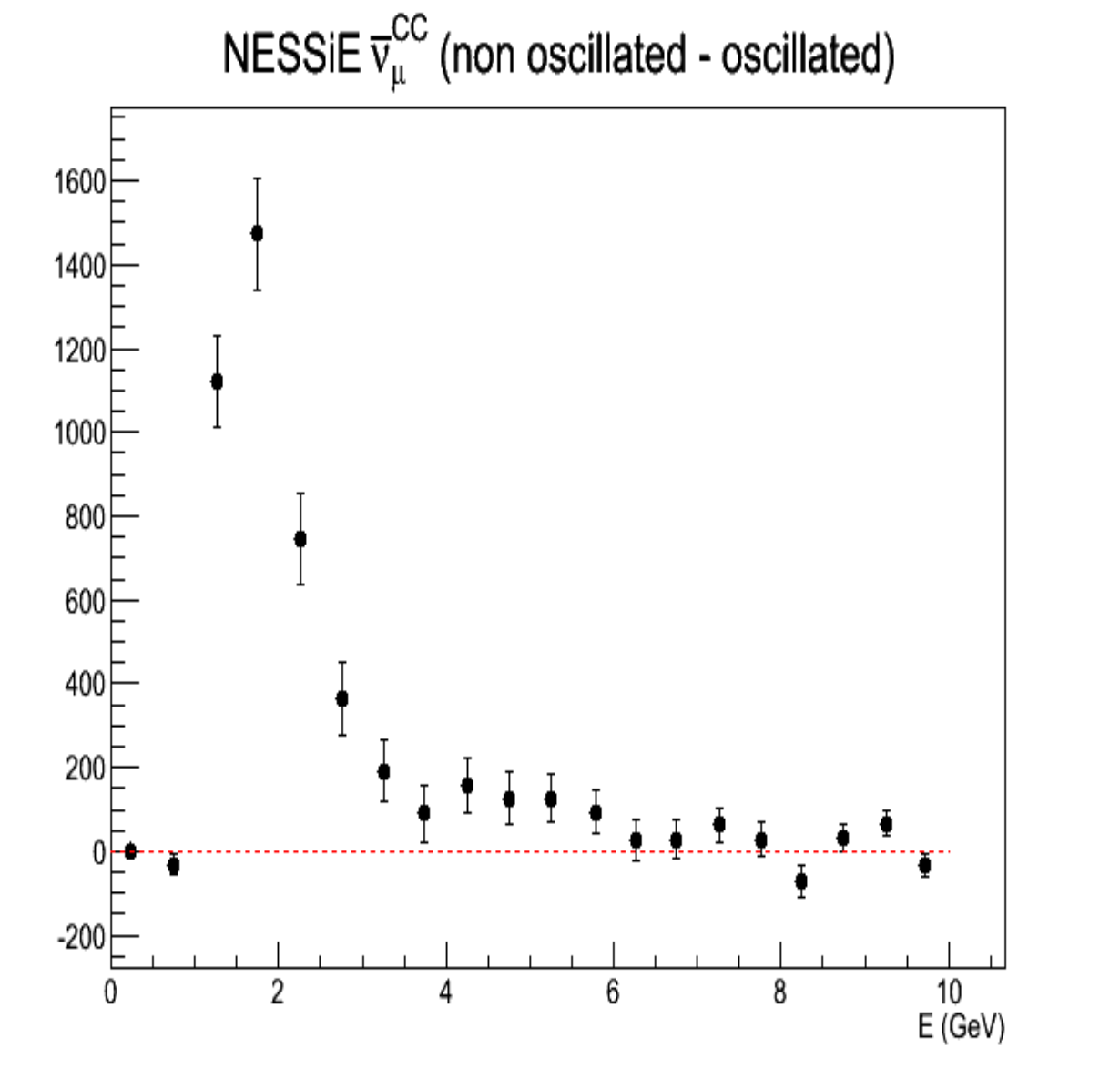}%
    \includegraphics[width=0.4\textwidth]{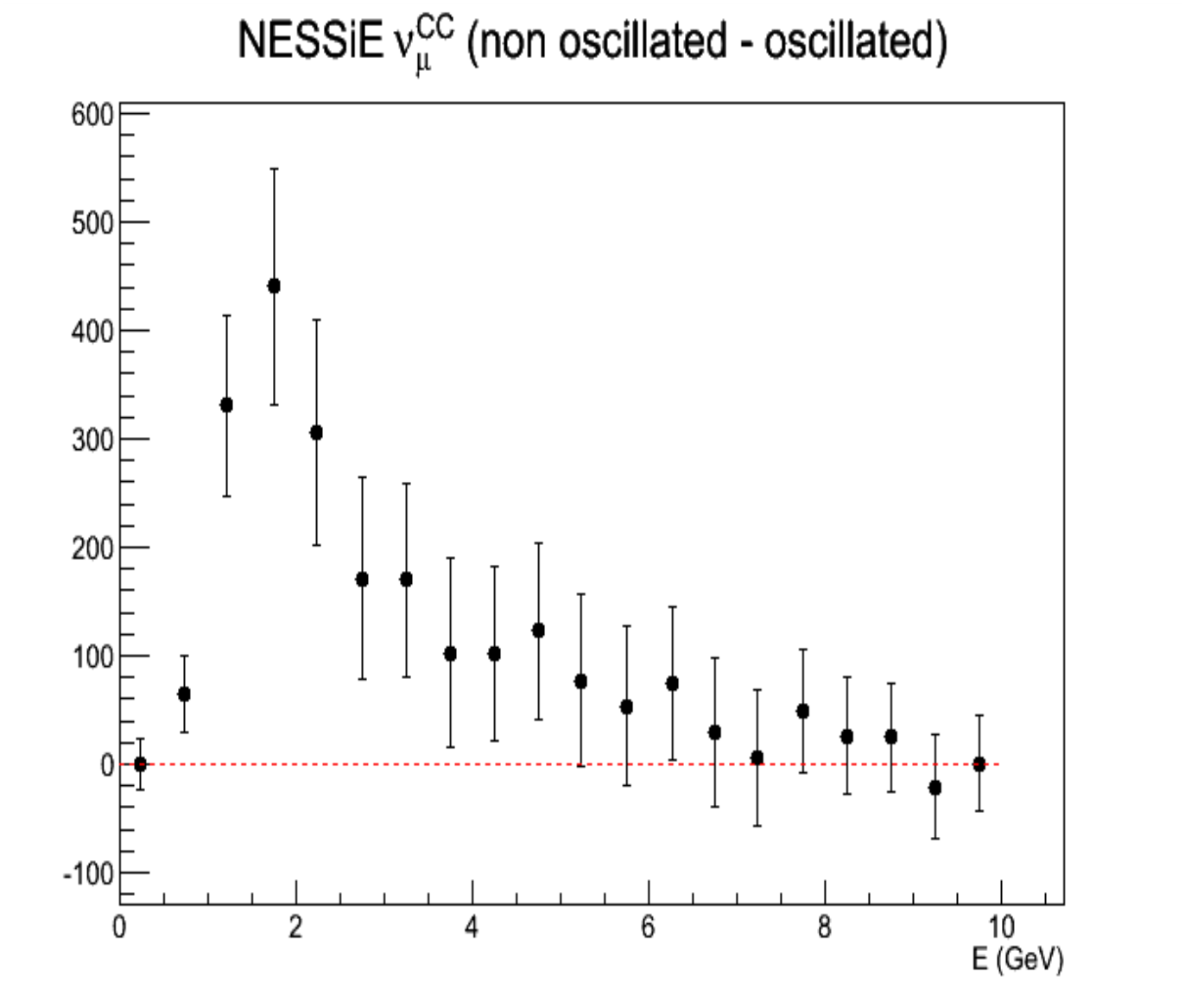}
    \caption{Disappearance signals as extracted from only one year of running in antineutrino mode. The intrinsic contribution of neutrinos
     in the running mode is sufficient to observe the possible disappearance signal in the neutrino channel, too. Only statistical errors are shown.}
    \label{disapp-test}
\end{center}
\end{figure}

The powerfulness of the statistical sample that can be collected in just one year of running at the proposed CERN neutrino beam is demonstrated
in Fig.~\ref{disapp-test}. Again for one year of running in negative focussing the disappearance signals are reported. In the negative focussing
running mode both the contributions of neutrinos and anti-neutrinos are sufficiently high to allow for measurements at the same time.

\begin{figure}[htbp]
\begin{center}
  \includegraphics[width=0.48\textwidth]{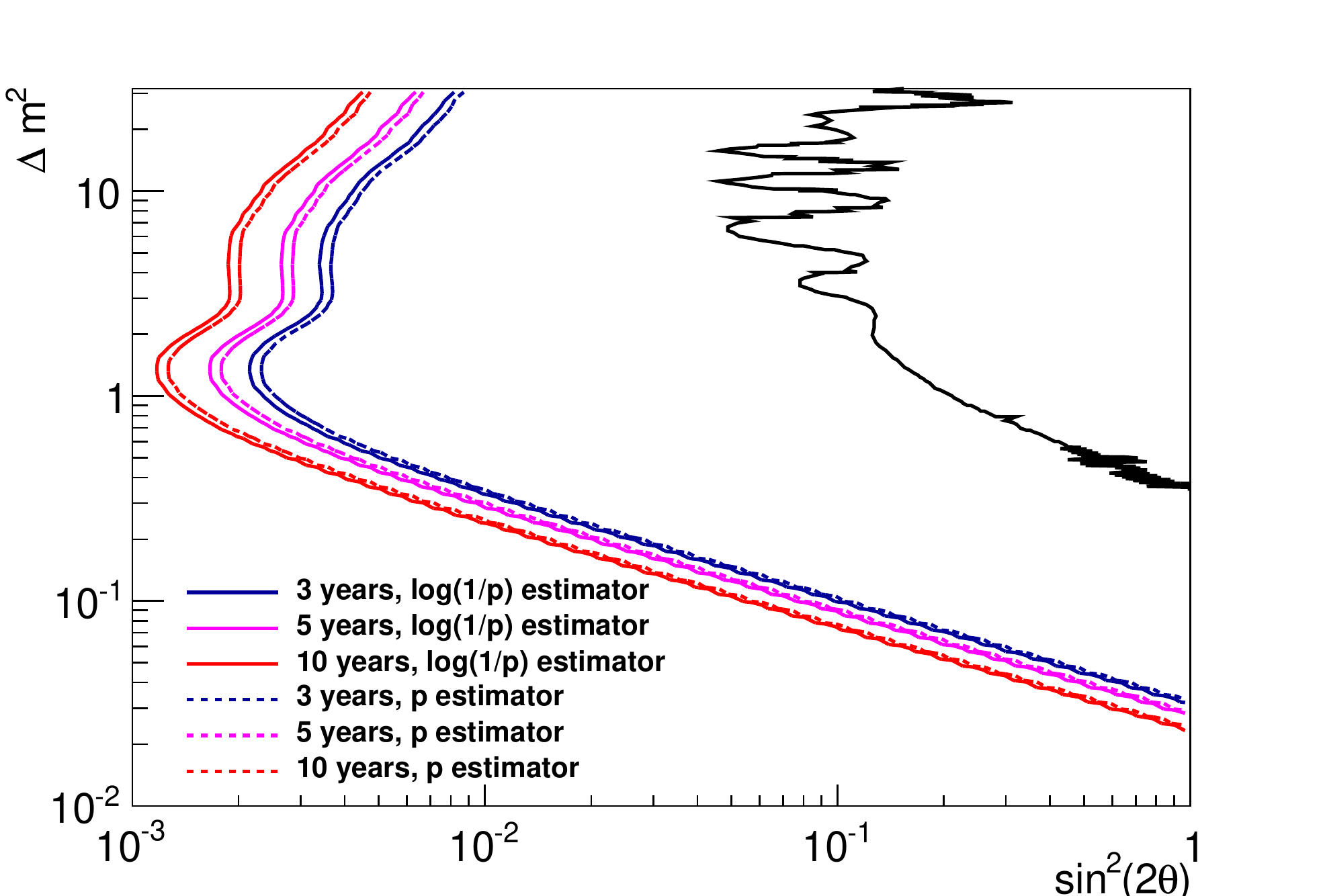}
   \includegraphics[width=0.48\textwidth]{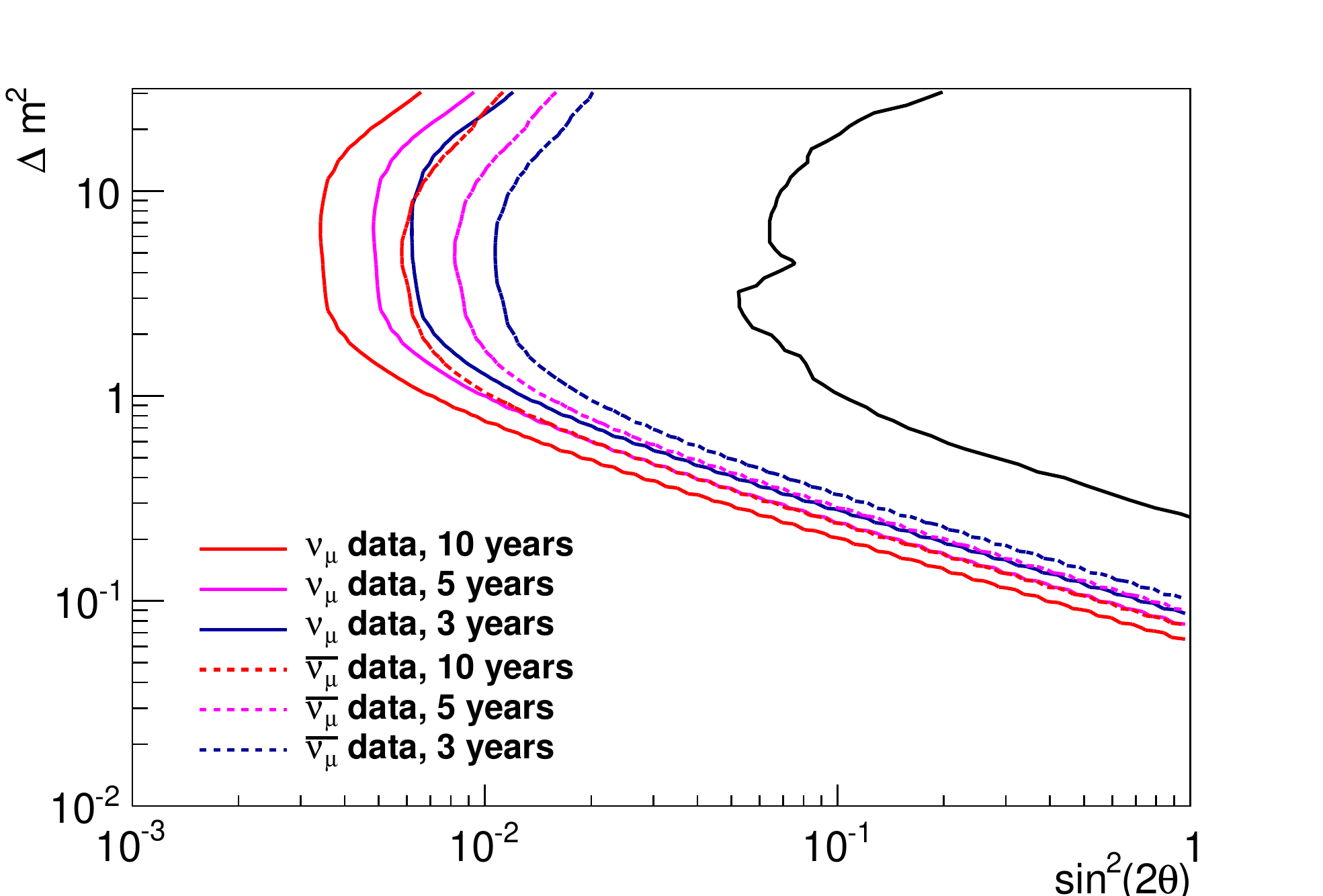}
    \caption{The estimated limits at 95\% C.L. (taken from ~\cite{stanco}) for $\nu_{\mu}$ disappearance at a Short--Baseline beam at CERN for several luminosity 
    running periods and different beam polarities,
    with a two--site massive spectrometer (770 tons and 330 tons, respectively) with 90\% inner fiducial volume.
    The left figure refers to the positive polarity beam. The continuous (dashed) lines correspond to the sensitivity limits obtained with the 
    $\log_{10}(1/p)$ ($p$) variable (used in ~\cite{stanco}). 3 years correspond to $13.5\cdot 10^{19}$ p.o.t., 5 years to $22.5\cdot 10^{19}$ p.o.t. and 10 years to $45.0\cdot 10^{19}$ p.o.t.
    The exclusion limit from combined MiniBooNE and SciBooNE $\nu_{\mu}$ disappearance result at 90\% C.L. from Ref.~\cite{mini-sci-mu1} is shown for comparison by the black curve in the right.
   The right figure refers to the negative polarity beam. Sensitivity limits are evaluated with the $\log_{10}(1/p)$ variable.
     Clearly the negative polarity run allows the contemporary analysis of the  $\overline\nu_{\mu}$ and $\nu_{\mu}$ disappearance exclusion regions 
     thanks to the disentangling of the muon charge on an event--by--event--basis. The black curve in the right shows for comparison the central value of the sensitivity at 90\% C.L.  from
     combined MiniBooNE and SciBooNE $\overline\nu_{\mu}$ disappearance result (Ref.~\cite{mini-sci-mu2}).}
    \label{ster-5}
\end{center}
\end{figure}

Finally, we outline in Fig.~\ref{ster-5} the estimation evaluated in the paper~\cite{stanco} (Fig.~7 of~\cite{stanco}).The estimated limits at 95\% C.L. on $\nu_{\mu}$ disappearance 
that can be achieved via the {\em Far/Near} estimator are shown for different data periods
(3, 5 and 10 years, corresponding to $13.5\cdot 10^{19}$, $22.5\cdot 10^{19}$ and $45.0\cdot 10^{19}$  p.o.t., respectively). The different results for $\nu_{\mu}$
and $\overline\nu_{\mu}$ beams were evaluated using the two variables, $p$ and $\log_{10}(1/p)$. In negative polarity runs the muon charge identification allows 
an independent, simultaneous and similar--sensitivity measurements of  the $\overline\nu_{\mu}$  and $\nu_{\mu}$ disappearance rates, due to the large $\nu_{\mu}$ 
contamination in the $\overline\nu_{\mu}$  beam. 

\section{Conclusions}

Neutrino physics is receiving more and more attention as a venue for the long standing search for
new physics beyond the Standard Model. The current anomalies which do not fit into the established standard scenario with 3 neutrinos
deserve refined studies and experiments. The CERN proposal for a new Short--Baseline experimental project is a very valuable one.
We illustrated the current critical tensions in the muon-neutrino disappearance field and the achievements that can be obtained
within the CERN project and the experiment proposed by the NESSiE Collaboration. Specifically, by considering $\mathcal{O}$(1) kton massive spectrometers,
 an improvement by an order of magnitude can be obtained in the sensitivity to the mixing parameter space between standard neutrinos
and sterile ones with respect to today's limits. Conversely, a possible $\nu_{\mu}$ disappearance signal will be essential to measure the
relevant physical parameters and to fully disentangle the different sterile models.
An effective analysis can be performed with a two-site experiment by using muon spectrometers with a low-$Z$ part that allows
clean charge identification on an event-by-event basis, and with a massive part allowing clean momentum  measurement through range.
Such a kind of spectrometer is under study by the NESSiE Collaboration and might be available with a limited investment using the OPERA spectrometers.
The performances of these kinds of detectors are suitable to put a
definitive result on the sterile neutrino issue at the eV mass scale.

\subsection*{Acknowledgements}

We warmly thank the Organizers of the Neutrino Telescopes Workshop for the kind invitation and the relevance they
endued to our project. We would also like to thank the CERN and INFN managements for their strong 
support towards the NESSiE Collaborations in view of the development of the foreseen CERN Neutrino-Platform.


\end{document}